\documentclass{aastex}          
\usepackage{spr-astr-addons}    
\usepackage{epsfig}
\usepackage{graphicx}
\usepackage{color}

\begin{document}
%
\title{Phase mixing of propagating Alfv\'{e}n waves in a stratified atmosphere: Solar spicules}

\shorttitle{Phase mixing of Alfv\'{e}n waves in solar spicules}
\shortauthors{Ebadi et al.}

\author{H.~Ebadi\altaffilmark{1}}
\affil{Astrophysics Department, Physics Faculty,
University of Tabriz, Tabriz, Iran\\
e-mail: \textcolor{blue}{hosseinebadi@tabrizu.ac.ir}}
\and
\author{M.~Hosseinpour}
\affil{Plasma Physics Department, Physics Faculty,
University of Tabriz, Tabriz, Iran}
\and
\author{H.~Altafi-Mehrabani}
\affil{Physics Faculty,
University of Zanjan, Zanjan, Iran}

\altaffiltext{1}{Research Institute for Astronomy and Astrophysics of Maragha,
Maragha 55134-441, Iran.}

\begin{abstract}
Alfv\'{e}nic waves are thought to play an important role in coronal heating and solar wind acceleration.
Recent observations by \emph{Hinode}/SOT showed that the spicules mostly exhibit upward propagating high frequency waves.
Here we investigate the dissipation of such waves due to phase mixing in stratified environment of solar spicules. Since
they are highly dynamic structures with speeds at about significant fractions of the Alfv\'{e}n phase speed,
we take into account the effects of steady flows. Our numerical
simulations show that in the presence of stratification due to gravity, damping takes place in space than in time.
The exponential damping low,~exp(-At$^{3}$), is valid under spicule conditions, however the calculated damping time
is much longer than the reported spicule lifetimes from observations.

\end{abstract}


\section{Introduction}
\label{sec:intro}
The coronal heating \citep{Edlen1943} mechanism is one of the major unsolved problems in solar physics.
Since the energy flux carried by acoustic waves is too small, the possibility of heating
by MHD waves has been investigated intensively as the magnetic structure of the solar corona
can play an important role here \citep{Hood1997}. The propagation of Alfv\'{e}n waves is one of the candidate
mechanisms that can carry energy to large distances from the surface, and heat the solar corona.
However, a heating theory based on the Alfv\'{e}n waves faces a couple of difficulties: Firstly, the waves have to
transport enough energy flux, and secondly, they have to dissipate efficiently in order to deposit the
right amount of energy at the right place \citep{De1999}.
The Alfv\'{e}n waves may reach the corona even in the absence of highly stratified atmosphere but with lesser propagation speed.
The damping length of Alfv\'{e}n modes is defined by various dissipative processes such as phase mixing \citep{Hey1983, Brown1991},
resonance absorption \citep{Ionson1978}, and nonlinear mode conversion \citep{Hollweg1982}. Phase mixing is a mechanism
for dissipating Alfv\'{e}n waves, which was first proposed by \citet{Hey1983}. When Alfv\'{e}n waves propagating
in an inhomogeneous medium, on each magnetic field line, a wave propagates with its own local Alfv\'{e}n speed.
After a certain distance or after enough time, these neighboring perturbations will be out of phase. This ultimately results in a strong
enhancement of the dissipation of Alfv\'{e}n waves energy via both viscosity and resistivity.
\citet{Hood2002} and \citet{Hey1983} analytically showed that in both the strong phase mixing limit and the weak
damping approximation, the amplitude of Alfv\'{e}n waves decays with time as ~exp(-t$^{3}$). \citet{Karami2009}
calculated numerically the damping times of oscillations in the presence of viscosity and resistivity in coronal loops. They concluded
that the above exponential damping law in time is valid for the Lundquist numbers higher than~$10^{7}$.
\citet{De1999, De2000} studied the effect of stratification due to gravity on phase mixing, and found that
the wavelengths lengthen when Alfv\'{e}n waves propagate through a stratified plasma. They concluded that a vertical
stratification of density makes phase mixing by ohmic heating less important in coronal heating problem.
They also found that in a stratified atmosphere, the heat will be deposited higher up than in an unstratified
atmosphere, and that the viscous heating will be the dominant component in the heating processes at lower heights.
However, in coronal conditions the effect of stratification on efficiency of phase mixing would still be large as
the height at which most heat would be deposited through ohmic dissipation is increased considerably by stratification.
Moreover, depending on the value of pressure scale height, phase mixing can either be more or less efficient than in the uniform
case.

Spicules have long been investigated as a coronal heating agent \citep{Hollweg1982, Athay1982}. They are grass-like spiky features
seen in chromospheric spectral lines at the solar limb \citep{Tem2009}. These spiky dynamic jets are propelled upwards
 at speeds of about $20-25$~km~s$^{-1}$ from photosphere into the magnetized low atmosphere of the sun. Their
diameter varies from spicule to spicule having the values from $400$~km to $1500$~km. The mean length of classical
spicules varies from $5000$~km to $9000$~km, and the typical life time of them is $5-15$ min.
The typical electron density at heights where the spicules are observed
is $3.5\times10^{10}-2\times10^{11}$ cm$^{-3}$, and their temperatures are estimated $\sim 5000-8000$ K \citep{bec68}. \citet{Kukh2006, Tem2007}
by analyzing the height series of $H\alpha$ spectra in solar limb spicules observed their transverse oscillations.
The period of them estimated $20-55$ and $75-110$ s. They concluded that these oscillations can be caused by propagating kink waves in spicules.
\citet{De2007} based on \emph{Hinode} observations concluded that the most expected periods of transverse oscillations lay between $100-500$ s, which
 interpreted as signatures of Alfv\'{e}n waves. \citet{Okamoto2011} used \emph{Hinode}/SOT observations
 of spicules, and concluded that upward propagating, downward propagating and standing waves occurred at the
 rates of about $60\% $, $20\%$ and $20\%$, respectively. Furthermore, they found that upward propagating waves dominate
 at lower latitudes, and the medians of amplitude, period, and velocity amplitude are $55$ km, $45$ s, and $7.4$ km s$^{-1}$,
 respectively. More recently \citet{Ebadi2012} made time-slice images of spicules, which were observed by \emph{Hinode}/SOT.
 They concluded that the energy flux stored in spicule axis oscillations is of the order of coronal energy
 loss in quiet sun. These results motivated us to study the phase mixing of upward propagating Alfv\'{e}n waves in a stratified
 atmosphere. To do so, the section $2$ gives the basic equations and the theoretical model. In section $3$ the numerical
 results are presented and discussed, and a brief summary is followed in section $4$.

\section{Theoretical modeling}
\label{sec:theory}

In the present work we keep the effects of stratification due to
gravity in $2$D x-z plane. The phase mixing and dissipation of propagating
Alfv\'{e}n waves in a region with non-uniform Alfv\'{e}n velocity is
studied. MHD equations governing the plasma dynamics are as follows:
\begin{equation}
\label{eq:cont} \frac{\partial \rho}{\partial t}+\nabla\cdot(\rho
\mathbf{v}) = 0,
\end{equation}
\begin{equation}
\label{eq:momentum} \rho\frac{\partial \mathbf{v}}{\partial t}+
\rho(\mathbf{v} \cdot \nabla)\mathbf{v} = -\nabla p + \rho
\mathbf{g}+ \frac{1}{\mu}(\nabla \times \mathbf{B})\times
\mathbf{B}+ \rho\nu\nabla^{2}\mathbf{v},
\end{equation}

\begin{equation}
\label{eq:induction} \frac{\partial \mathbf{B}}{\partial t} = \nabla
\times(\mathbf{v} \times \mathbf{B})+ \eta\nabla^{2}\mathbf{B},
\end{equation}

\begin{equation}
\label{eq:state} p = \frac{\rho RT}{\mu},
\end{equation}
\begin{equation}
\label{eq:divB} \nabla\cdot \mathbf{B} = 0,
\end{equation}
where $\nu$ and $\eta$ are constant viscosity and resistivity coefficients,
and other quantities have the usual meaning. In
particular, typical values for $\eta$ in the solar chromosphere and corona are
$8\times10^{8}T^{-3/2}$ and $10^{9}T^{-3/2}$ m$^{2}$s$^{-1}$, respectively. The
value of $\rho\nu$ for a fully ionized H plasma is $2.2\times10^{-17}T^{5/2}$
kg m$^{-1}$ s$^{-1}$ \citep{Priest1982}.
We assume that the spicules are highly dynamic with speeds that are
significant fractions of the Alfv\'{e}n speed. The perturbations are assumed independent of y,
with a polarization in \^{y} direction, i.e.:
\begin{eqnarray}
\label{eq:perv}
  \textbf{v} &=& v_{0} \hat{k} + v_{y}(x,z,t) \hat{j} \nonumber\\
  \textbf{B} &=& B_{0} \hat{k} + b_{y}(x,z,t) \hat{j}.
\end{eqnarray}

Therefore, the pressure gradient is balanced by the gravity force, which is assumed
to be \textbf{g}=g $\hat{k}$ via this equation:

\begin{equation}
\label{eq:balance}
 -\nabla p_{0} + \rho_{0} \textbf{g}=0,
\end{equation}

and the pressure in an equilibrium state is:
\begin{equation}
\label{eq:press}
 p_{0}= p_{0}(x)e^{-z/H}.
\end{equation}

\begin{figure}[!h]
\epsscale{1.0} \plotone{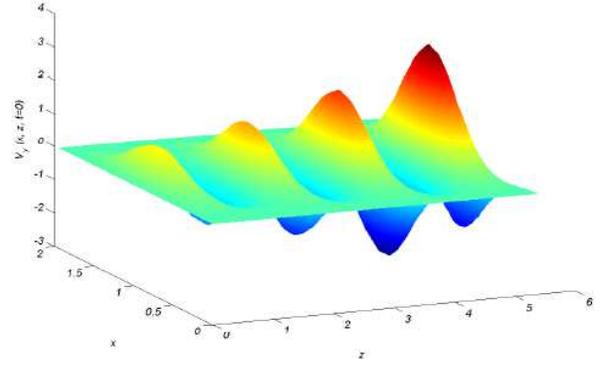} \caption{ The plot of initial wave packet with d=0.3a is presented. \label{fig1}}
\end{figure}

The density profile is in the form of:

\begin{equation}
\label{eq:density}
 \rho_{0}= \rho_{0}(x)e^{-z/H},
\end{equation}
with
\begin{equation}
\label{eq:scale}
 H= \frac{RT}{\mu g},
\end{equation}
where $H$ is the pressure scale height.
The linearized dimensionless MHD equations with these assumptions are:

\begin{equation}
\label{eq:velo}
 \frac{\partial v_{y}}{\partial t} + v_{0}\frac{\partial v_{y}}{\partial z}
 = V^{2}_{A}(x,z)\frac{\partial b_{y}}{\partial z}+ \nu\nabla^{2}v_{y},
\end{equation}

and

\begin{equation}
\label{eq:mag}
 \frac{\partial b_{y}}{\partial t} + v_{0}\frac{\partial b_{y}}{\partial z}
 = \frac{\partial v_{y}}{\partial z}+ \eta\nabla^{2}b_{y},
\end{equation}
where the velocities, the magnetic field, time and space coordinates are normalized to V$_{A0}\equiv B_{\rm 0}/\sqrt{\mu \rho_{\rm 0}}$
(with $\rho_{\rm 0}$ as the plasma density at $z=0$),
$B_{\rm 0}$, $t_{A}$ (the period of Alfv\'{e}n waves), $a$ (spicule radios), respectively. Also the resistivity and viscosity coefficients
are normalized to $a^{2}/t_{A}$.
The second terms in the left hand side of eqs.~\ref{eq:velo}, and~\ref{eq:mag} present the effect of steady flows.
$V_{\rm A}(x,z)$ is the Alfv\'{e}n velocity, which for
a phase mixed and stratified atmosphere due to gravity is
assumed to be \citep{De1999,Karami2009}:

\begin{equation}
\label{eq:av}
 V_{A}(x,z)=V_{A0}e^{z/2H}[2+\tanh[\alpha(x-1)]],
\end{equation}
where parameter $\alpha$ controls the size of inhomogeneity across the magnetic field. \\
The set of eqs.~\ref{eq:velo}, and~\ref{eq:mag} should be solved under these
initial and boundary conditions:

\begin{eqnarray}
\label{eq:icv}
  v_{y}(x,z,t=0) &=& V_{A0}\exp \left [-\frac{1}{2}(\frac{x-1}{d})^{2}\right]\sin(kz)e^{z/4H} \nonumber\\
  b_{y}(x,z,t=0) &=& A \sin(\pi x)\sin(\pi z) ,
\end{eqnarray}
where d is the width of the initial packet and $A=10^{-7}$. Figure~\ref{fig1} is the plot of initial wave packet given by
equation~\ref{eq:icv} for $d=0.3a$ ($a$ is the spicule radius).
The parameter $k$ is chosen in such a way to have upward propagating Alfv\'{e}n wave.

\begin{eqnarray}
\label{eq:icx}
  v_{y}(x=0,z,t)=v_{y}(x=2,z,t)=0 \nonumber\\
  b_{y}(x=0,z,t)=b_{y}(x=2,z,t)=0,
\end{eqnarray}

\begin{equation}
\label{eq:icd}
 \frac{\partial v_{y}}{\partial t}\bigg|_{t=0}=0.
\end{equation}

\section{Numerical results and discussion}

To solve the coupled eqs.~\ref{eq:velo}, and~\ref{eq:mag} numerically,
the finite difference and the Fourth-order Runge-Kutta methods
are used to take the space and time derivatives, respectively.
We set the number of mesh grid points as~$256\times256$.
In addition, the time step is chosen as $0.002$ (time is normalized to the Alfv\'{e}n
time, $t_{A}$), and the system length in the $x$ and $z$ dimensions
(simulation box sizes) are set to be $2000$ km and $6000$ km.
The parameters in spicule environment are as follows \citep{Tem2010,Ebadi2012}:
$a$ (spicule radios)=500 km,
$d=0.3a=150 km$ (the width of gaussian paket), L=6000 km (Spicule length), $v_{0}=25 km/s$, $B_{0}=10 G$, $n_{e}=10^{11} cm^{-3}$,
 $T=8000 K$, $g=272 m s^{-2}$, $R=8300 m^{2}s^{-1}k^{-1}$ (universal gas constant),
$V_{A0}=40 km/s$,  $k=4\pi/3$ (dimensionless wavenumber normalized to $a$), $\nu=10^{3} m^{2}s^{-1}$, $\eta=10^{3} m^{2}s^{-1}$,
 $\mu=0.6$,  $H=500 km$, $t_{A}=37.5$ s (the period of Alfv\'{e}n waves), and $\alpha=2$ \citep{Okamoto2011}.\\

\begin{figure}[!h]
\epsscale{1.0} \plotone{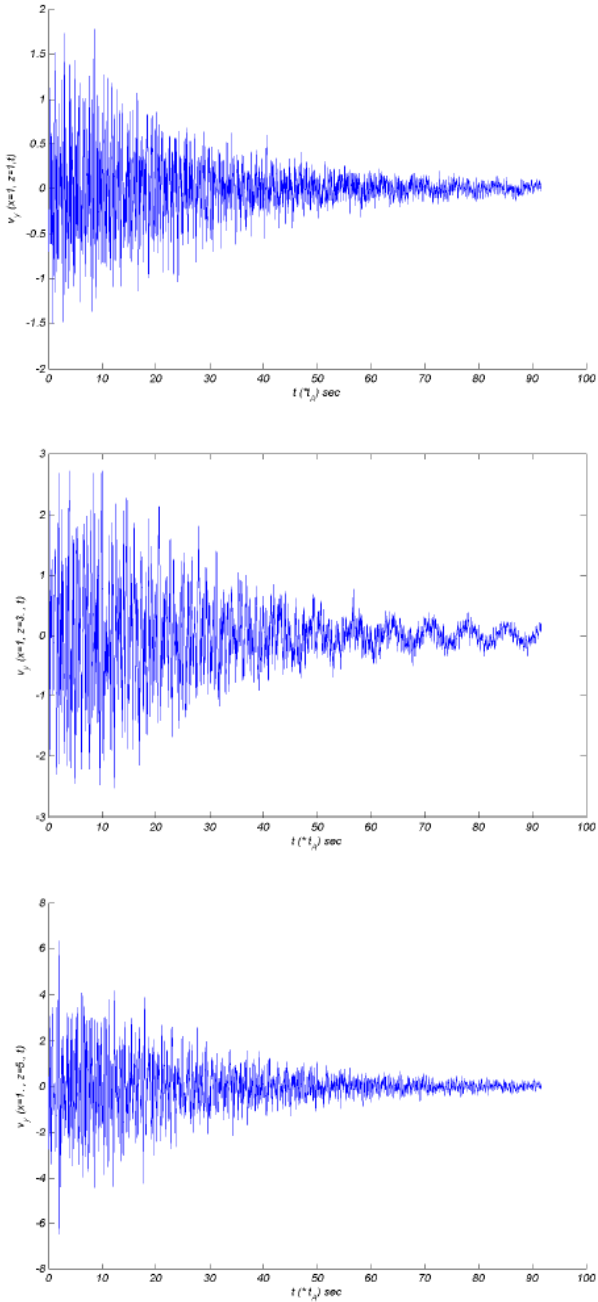} \caption{ The perturbed velocity variations with respect to time
in $x=1000$ km, $z=1000$ km; $x=1000$ km, $z=3000$ km; and $x=1000$ km,
$z=5000$ km respectively from top to bottom are showed.\label{fig2}}
\end{figure}

\begin{figure}
\centering
\includegraphics[width=8cm]{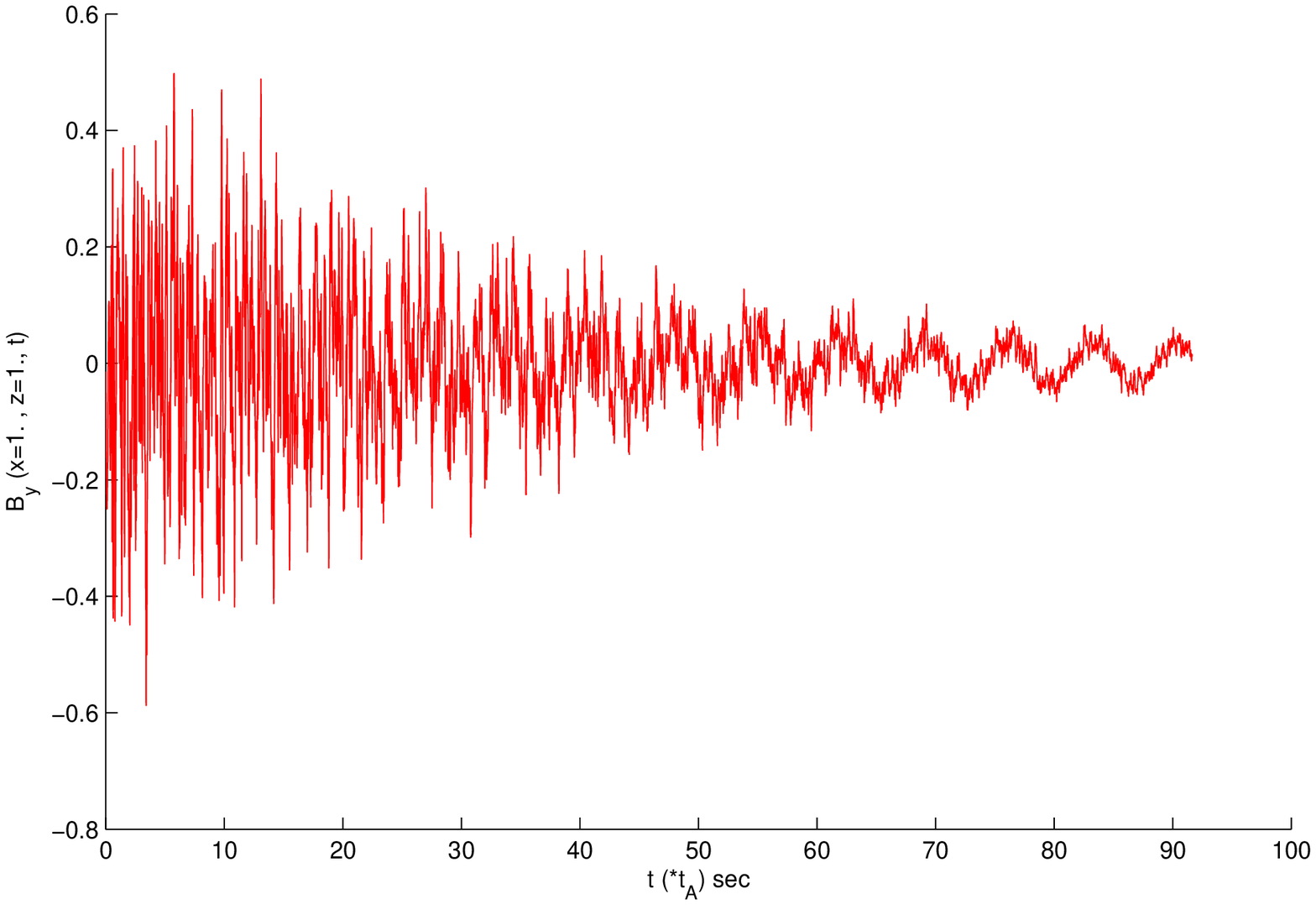}
\includegraphics[width=8cm]{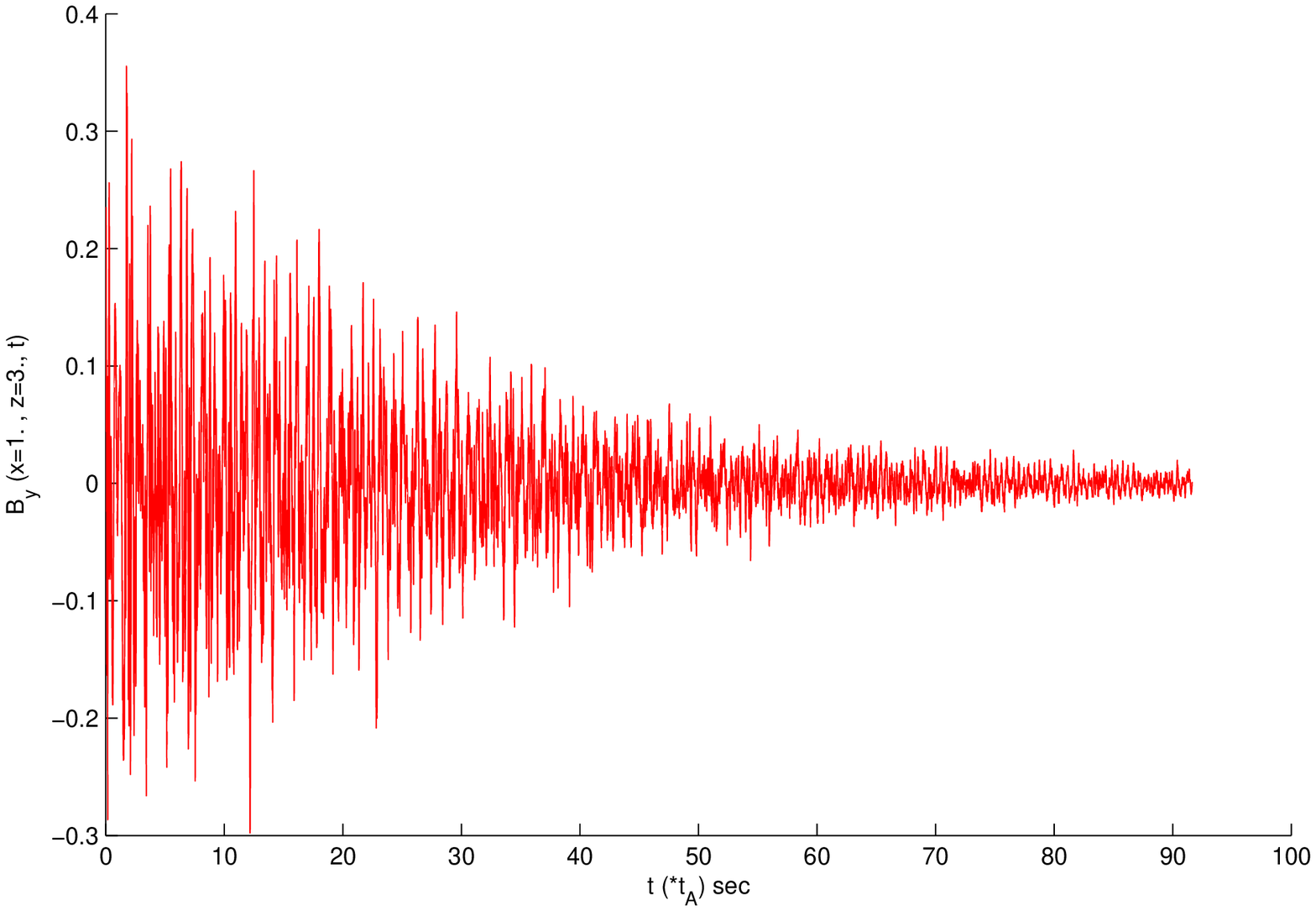}
\includegraphics[width=8cm]{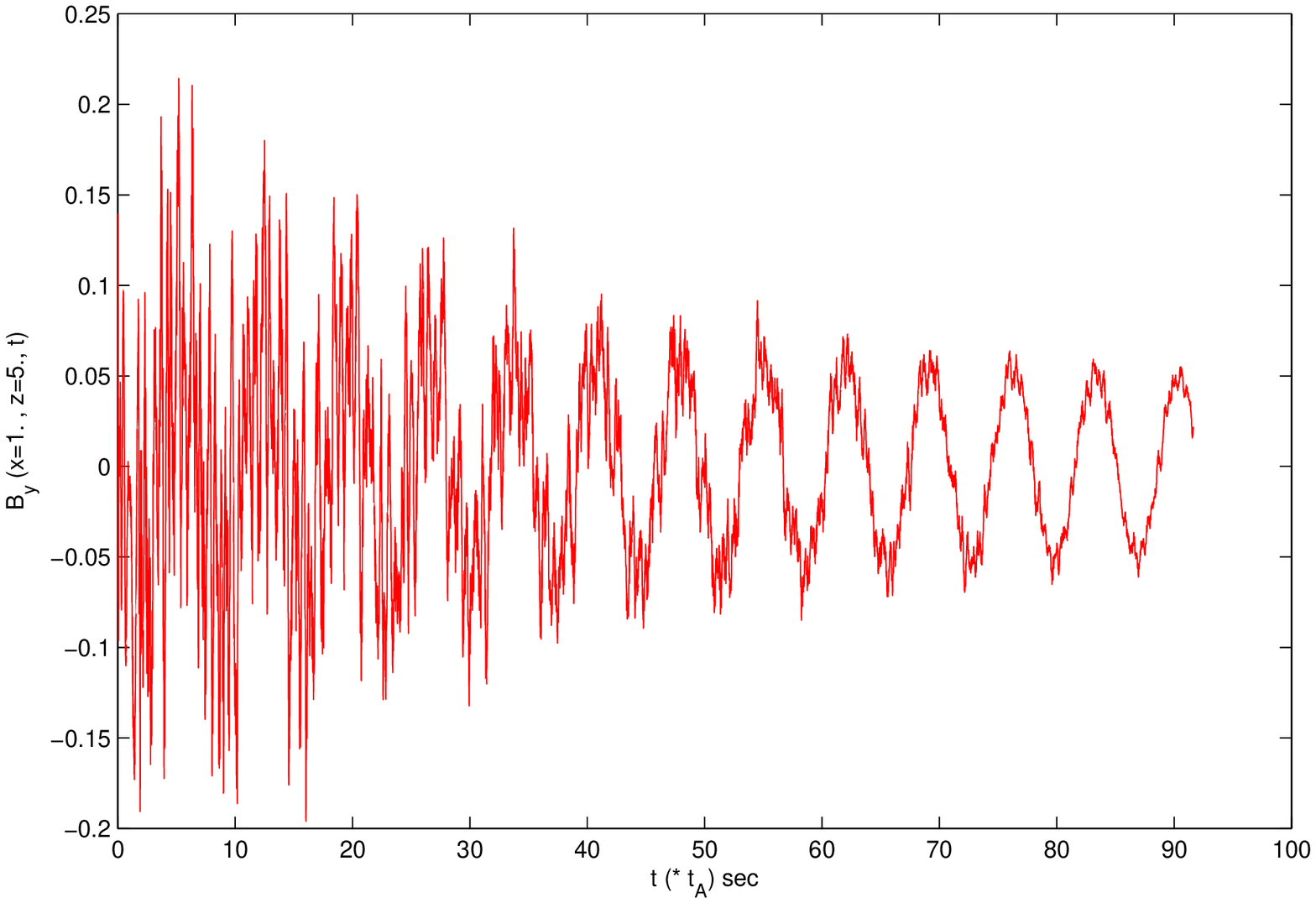}
\caption{The same as in Figure~\ref{fig2} but for the perturbed magnetic field.\label{fig3}}
\end{figure}

Figure~\ref{fig2} shows the perturbed velocity variations with respect to time
in $x=1000$ km, $z=1000$ km; $x=1000$ km, $z=3000$ km; and $x=1000$ km, $z=5000$ km respectively. We presented
the perturbed magnetic field variations obtained from our numerical analysis in Figure~\ref{fig3}
for $x=1000$ km, $z=1000$ km; $x=1000$ km, $z=3000$ km; and $x=1000$ km, $z=5000$ km respectively.
In these plots the perturbed velocity and
magnetic field is normalized to $V_{A0}$ and $B_{0}$ respectively.
In each set of plots it is appeared that both the perturbed
velocity and magnetic field are damped in the developed stage of phase mixing.

%

%

%
Further, at the first height (1000 km), total amplitude of both velocity and magnetic
field oscillations have values near to the initial ones. As height increases,
the perturbed velocity amplitude does increase in contrast to the behavior of perturbed magnetic field.
Nonetheless, exponentially damping behavior
is obvious in both cases. This means that with an increase in height,
amplitude of velocity oscillations is expanded due to significant decrease
in density, which acts as inertia against oscillations. Similar results are observed
by time-distance analysis of Solar spicule oscillations \citep{Ebadi2012}.
It is worth to note that the density stratification influence on the magnetic field is negligible,
which is in agreement with Solar Optical Telescope observations of Solar spicules \citep{Verth2011}.

\begin{figure}[!h]
\epsscale{1.0} \plotone{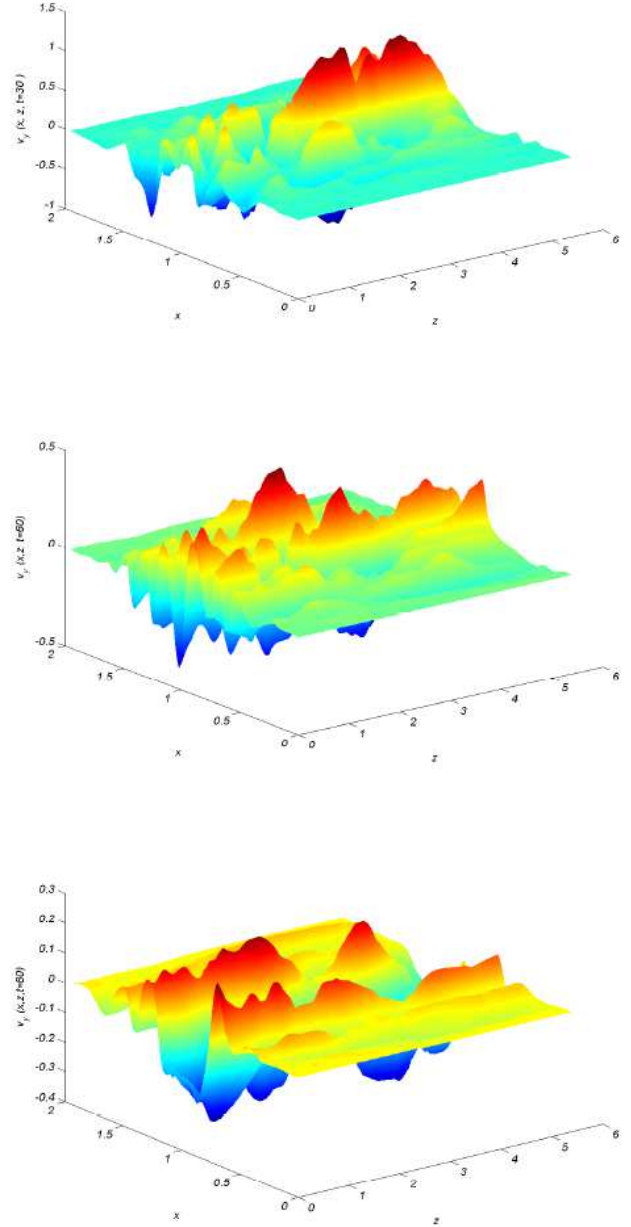} \caption{ (Color online) The perturbed velocity in $x-z$ space is presented. The panels from top to bottom correspond
to $t=30 t_{A}$, $t=60 t_{A}$, and $t=80 t_{A}$, respectively.    \label{fig4}}
\end{figure}

\begin{figure}[!h]
\epsscale{1.0} \plotone{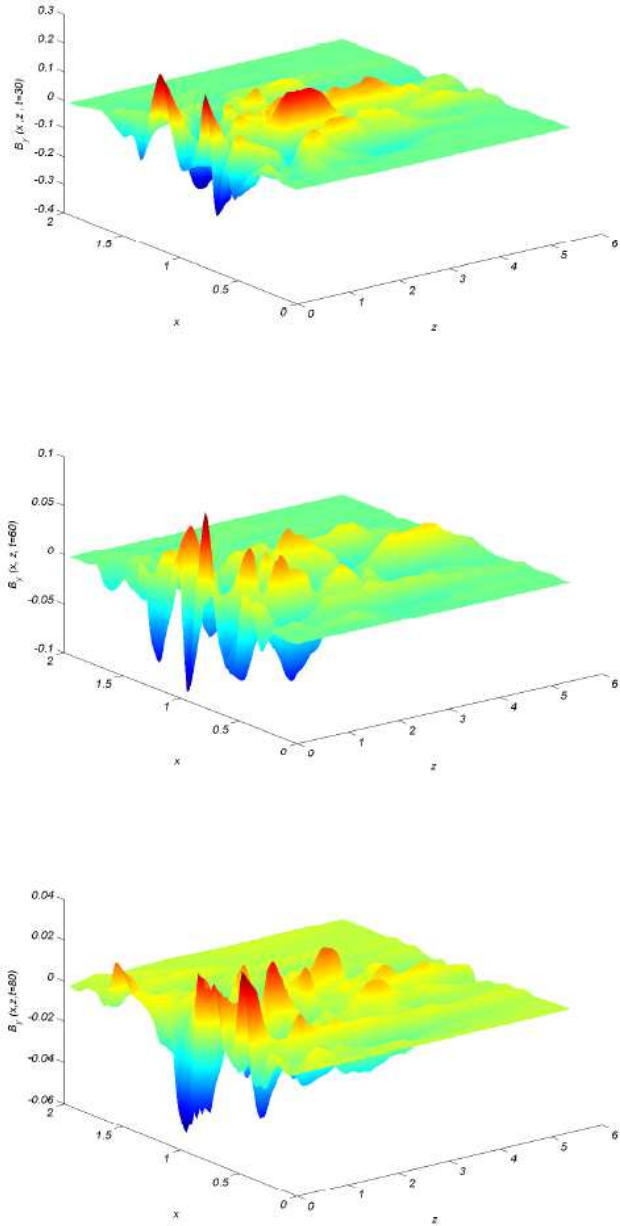} \caption{ (Color online) The same as in Figure~\ref{fig4} but for the perturbed magnetic field.\label{fig5}}
\end{figure}

Figures~\ref{fig4},~\ref{fig5} show the $3D$ plots of the perturbed velocity and magnetic field with respect to
$x$, $z$ for $t=30 t_{A}$, $t=60 t_{A}$, and $t=80 t_{A}$. They show that in the presence of stratification due to gravity,
the damping takes place in space than in time as an important point of these graphs.
It should be emphasized that the damping time scale of the velocity field pattern is
longer than the corresponding magnetic field pattern due to the initial conditions.
In other words, in spite of standing waves, propagating waves
are stable and dissipate after some periods due to phase mixing \citep{Hey1983}.
%
%
%
%
%

For calculating the damping time it is suitable to calculate the total energy (kinetic energy plus magnetic energy)
per unit of length in $y$ direction as:

\begin{equation}
\label{eq:tenergy}
 E_{tot}(t) = \frac{16\pi}{B_{0}^{2}aL} \int_{0}^{2} dx \int_{0}^{6} dz [\rho (x,z) v_{y}^{2}(x,z) + b_{y}^{2}(x,z)].
\end{equation}

\begin{figure}[!h]
\epsscale{0.90} \plotone{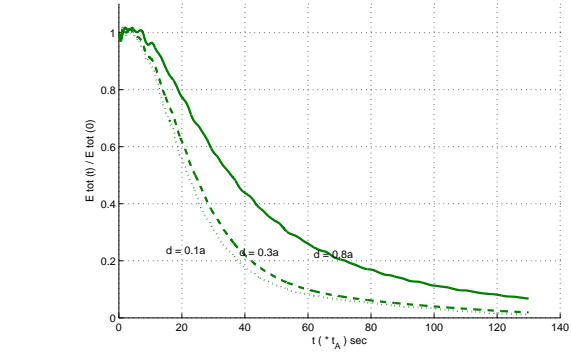}
\caption{The normalized total energy for three different initial wave packet widths is presented.
It is normalized to its value at $t=0$. \label{fig6}}
\end{figure}

In Figure~\ref{fig6} we plot the normalized total energy for three initial wave packet widths,
i.e. $d=0.1a$, $d=0.3a$, and $d=0.8a$.

\begin{figure}
\centering
\includegraphics[width=8cm]{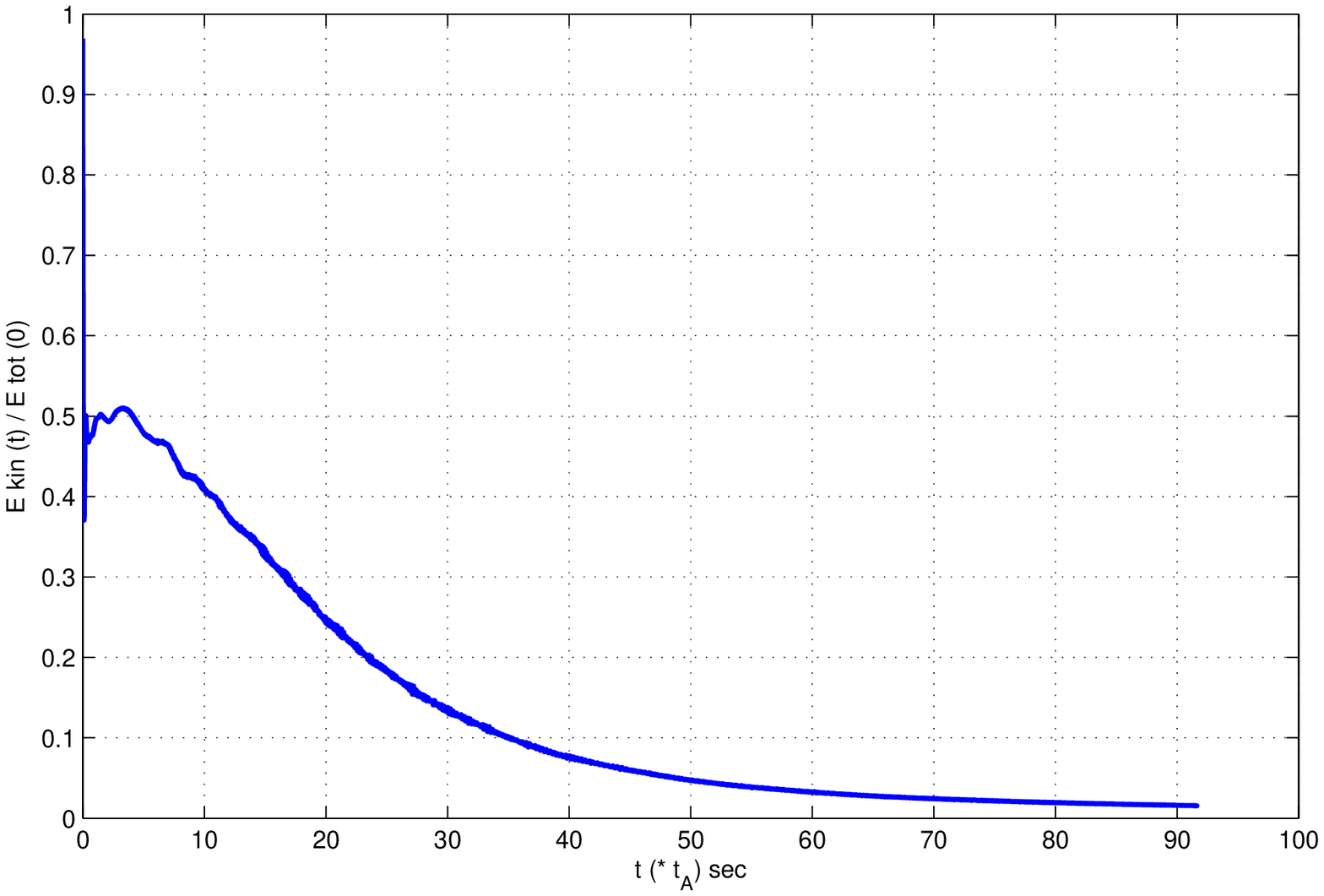}
\includegraphics[width=8cm]{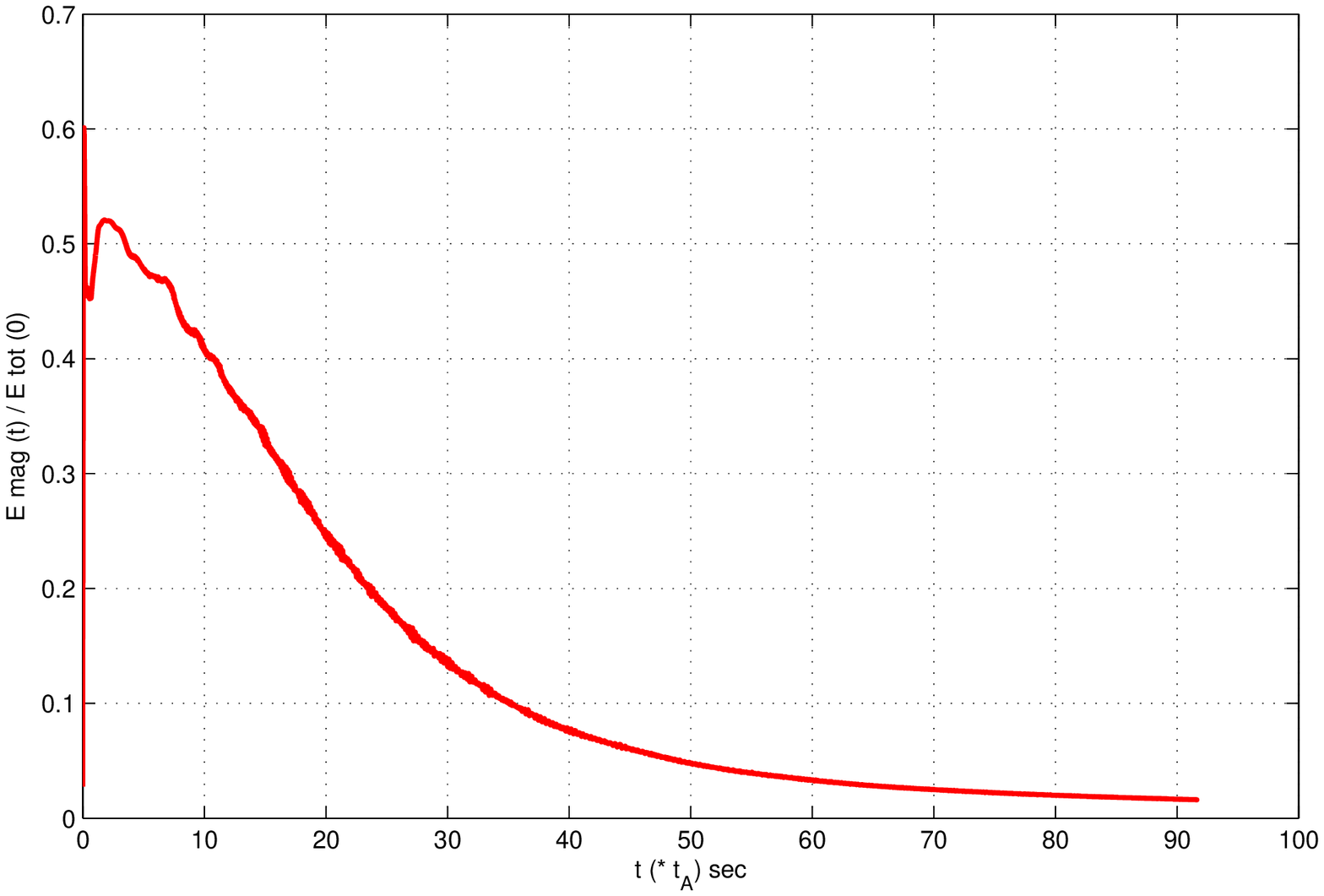}
\includegraphics[width=8cm]{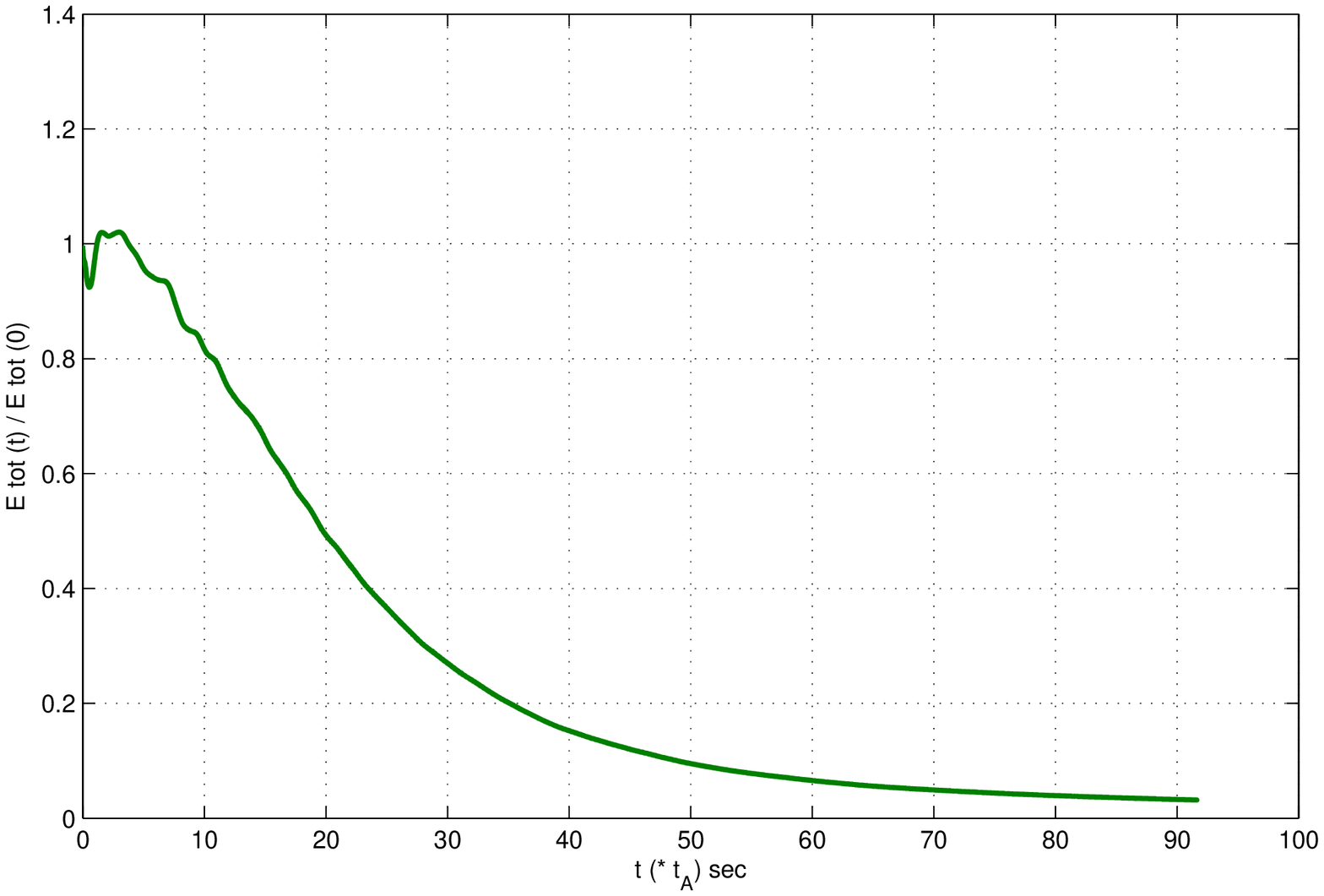}
\caption{Time variations of normalized kinetic energy, magnetic energy, and total energy for $d=0.3a$
are presented from top to bottom respectively.\label{fig7}}
\end{figure}

Since the treatments of different profiles of total energy are similar and the damping times are very close to each other,
hence, we continue our calculations with $d=0.3a$, which is more logical compared with tube radius.
In Figure~\ref{fig7} the kinetic energy, magnetic energy, and total energy
normalized to the initial total energy are presented respectively from top to bottom.
The damping time calculated from total energy profile is $1050$ s.
Actually, spicules have short lifetimes, and are transient phenomena.
We claim that in such circumstances, phase mixing can occur
in space rather than in time. This is in agreement with previous works \citep{De1999}.
We can definitely consider it as an uncertain rule of thumb.

The total energy profile is best fitted to the exponential damping function in time,~exp(-At$^{B}$), with $A=10^{-9}$, $B=2.85$.
Since the spicules are structures with low resistivity and viscosity coefficients, so this is in agreement with \citet{Hey1983} work.

So far the presented results in this paper are taken with considering a steady flow ($v_{0}=25 km/s$).
Actually, it is of interest to investigate the differences in our numerical analysis outputs if the flow effect is omitted.
To do that, we repeated the numerical analysis with zero-steady flow.
Interestingly, the relevant results did not
show significant difference in respect to the case of non-zero steady flow.
A short theoretical scaling may help to clarify this result; comparing two
dimensionless terms in eq.~\ref{eq:velo}: the convective term, which is
associated with the steady flow, $v_{0}\frac{\partial v_{y}}{\partial z}$, and
the Lorentz term, $V^{2}_{A}(x,z)\frac{\partial b_{y}}{\partial z}$, yields:

\begin{eqnarray}
\label{eq:com}
  (V^{2}_{A}(x,z)\frac{\partial b_{y}}{\partial z})/(v_{0}\frac{\partial v_{y}}{\partial z})&\sim& (V^{2}_{A0}e^{z/H}\frac{\bar{b}_{y}}{L})/\frac{\bar{v}_{y}}{L} \nonumber\\
  &\sim& e^{z/H}\bar{b}_{y}/\bar{v}_{y} ,
\end{eqnarray}

Here $\bar{b}_{y}$, $\bar{v}_{y}$ are the average values of $b_{y}$, $v_{y}$ normalized to $B_{0}$ and $V_{A0}$.
Also, $v_{0}$ is normalized to $V_{A0}$ assuming that both have the same order of magnitudes. Moreover, $V_{A}\thickapprox V_{A0} e^{z/(2H)}$
has been taken into account. Now with the large values of $z$,
and assuming that $\bar{b}_{y}$ and $\bar{v}_{y}$ have a same order of
magnitudes, then $\frac{e^{z/H}\bar{b}_{y}}{\bar{v}_{y}}\gg1$. Meaning that the effect of the Lorentz term which is
associated with the stratification effect is more important than the dominated convective term
bringing into account the steady flow effects. Thus as long as $z/H\gg1$, one might ignore the steady flow effects here.
Similar scaling can be done in eq.~\ref{eq:mag} to show that in the presence of stratification
the steady flow effects are ignorable.

\section{conclusion}
\label{sec:concl}

In our simple model, we assume that spicules are small scale structures
(relative to coronal loops and other mega structures) with a uniform magnetic
field, and a uniform temperature in all heights, and the transition region
between chromosphere and corona has been neglected.
Density change along spicule axis is considerable, and stratification due
to gravity is significant. As a result, the medium is dense in its lower heights,
but it becomes rare and rare as height increases. Also, spicules have short
lifetimes, and are transient phenomena. We claim that in such circumstances,
phase mixing can occur in space rather than in time.
This is in agreement with previous studies.
In ordinary heights of a spicule, there is enough space to amplitude of a
wave to damp, and to transfer its energy to the medium. Furthermore, we observed that
the amplitude of perturbed velocity field, magnetic field and total energy
decreases with height exponentially. Our numerical analysis show that the main phase of
evolution dynamics occurs only in the first one third of the spicule height.
Furthermore, we repeated the numerical analysis with zero-steady flow. Interestingly, the relevant results did not
show significant difference in respect to the case of non-zero steady flow.

\acknowledgments
The authors thank Prof. De Pontieu for his useful comments in preparing this manuscript.
This work has been supported financially by Research Institute for Astronomy and Astrophysics of Maragha (RIAAM), Maragha, Iran.

\makeatletter
\let\clear@thebibliography@page=\relax
\makeatother

\end{document}